\documentclass[12 pt,reqno]{revtex4}
\usepackage{epsfig}
\usepackage{amssymb}

\newcommand{\be}{\begin{equation}}
\newcommand{\ee}{\end{equation}}
\newcommand{\ba}{\begin{eqnarray}}
\newcommand{\ea}{\end{eqnarray}}
\usepackage[T1]{fontenc}
\usepackage[cp1250]{inputenc}
\begin{document}
\title{Two-electron resonances in quasi-one dimensional quantum dots with Gaussian confinement}

\author{Arkadiusz Kuro\'s, Anna Okopi\'nska \\
Institute of Physics,  Jan Kochanowski University\\
ul. \'Swi\c{e}tokrzyska 15, 25-406 Kielce, Poland}

\begin{abstract}
We consider a quasi one-dimensional quantum dot composed of two Coulombically interacting electrons confined in a Gaussian trap. Apart from bound states, the system exhibits resonances that are related to the autoionization process. Employing the complex-coordinate rotation method, we determine the resonance widths and energies and discuss their dependence on the longitudinal confinement potential and the lateral radius of the quantum dot.
The stability properties of the system are discussed.
\end{abstract}
\maketitle
\section{Introduction}

Recently, it has become possible to fabricate few-particle systems that realize simple models of quantum theory and enable quantitative comparison with the accurate solutions of the Schr\"{o}dinger equation. The advantage of the produced nanosystems composed of a few isolated atoms~\cite{jochScience} or ions~\cite{ion}, as well as larger systems with a few-particle substructure such as semiconductor quantum dots~\cite{QD} is that their parameters can be experimentally controlled. Not only the number of constituents, but also the interactions between them and the geometry of the system can be modelled at will by applying appropriately designed electromagnetic fields. Those systems create a versatile platform for testing the effectiveness of approximation methods used in solving quantum many-body problems. Particularly fortunate from the point of view of comparison with theoretical considerations are the quasi-one-dimensional systems for which accurate few-body calculations are possible.   

The new experimental possibilities gave an impetus for accurate theoretical studies of simple two-body systems subjected to external potentials. Many theoretical works have discussed the properties of bound states of two Coulombically interacting particles confined by a harmonic potential~\cite{zw,1Da,xie}, much less studies were devoted to systems that show a reso\-nant behavior. In order to determine the resonance energy and  lifetime, various theoretical approaches have been used based on bound-state methods, e.g. the complex coordinate method \cite{by,mo,ge,ho}, the box approach \cite{box}, the complex absorbing potential  \cite{cap} and the real stabilization method \cite{os,ho2}. Investigation of autoionizing resonant states in two-particle systems has been performed for 3D atomic systems, helium and helium-like ions \cite{ho,ho2}, and spherically symmetric quantum dots~\cite{by,mo,ge}. The presence of autoionizing states is highly important for transport phenomena in nanosystems. The role of resonances in the scattering process in one-dimensional quantum dots has been also investigated \cite{1D}.

In this paper, we consider a system of two Coulombically interacting particles that are strongly confined laterally and weakly confined by the longitudinal potential which supports both bound and continuum stationary states. The system is modelled by a quasi-one-dimensional Hamiltonian with the parameters describing the shape of the confining potential and the interparticle interaction strength. We discuss the energy spectrum and study how the presence of autoionizing resonances depends on the system parameters. In particular, studying the dependence of the lifetime of the resonant state on the lateral confinement range, we will establish its influence on the stability properties of the quantum dot.

\section{The model}
\begin{figure}[h]
\begin{center}
\includegraphics[width=0.65\textwidth]{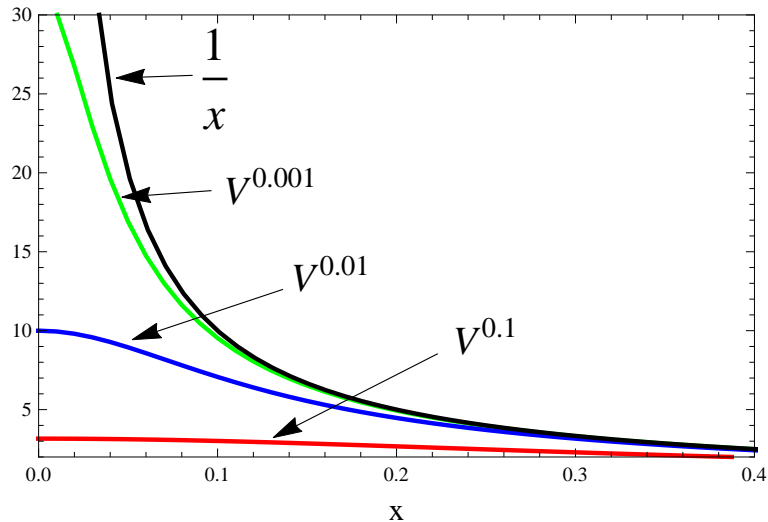}
\end{center}
\caption{The behavior of the truncated Coulomb interaction (\ref{pot}) for different $\delta$ parameters compared to the Coulomb bare potential where $x=|x_1-x_2|$.}
\label{ryspot}
\end{figure}

Our model approximates a  two-electron system in an axially symmetric anisotropic trap, where the lateral confinement is much stronger than the longitudinal one, so that the assumption that all excitations occur only in the longitudinal direction is justified and the system can be effectively described by a quasi-one-dimensional Hamiltonian 
\begin{equation}
\hat{H}=\sum_{i=1}^{2}\left[-\frac{1}{2}\frac{\partial^2}{\partial
x_i^2}-V_0 e^{- x_i^2}\right] + V^{\delta}(|x_1-x_2|).
 \label{ham}
\end{equation}
The effective interaction potential is taken in the truncated Coulomb form  \cite{1DCoulomb}
\begin{equation}
V^{\delta}(|x_1-x_2|)=\frac{g}{\sqrt{(x_1-x_2)^2+\delta}},
 \label{pot}
\end{equation}
where $g$ is the strength of the interaction. The parameter $\delta$ is related to the lateral confinement range which determines the lateral radius of the quasi-one-dimensional quantum dot. The simplified form~(\ref{pot}) has the same behavior at large interparticle distances as the bare Coulomb potential (see Fig. \ref{ryspot}) and is convenient to apply in numerical calculation based on exact diagonalization of the Hamiltonian. The limit of $\delta\rightarrow 0$ corresponds to the strictly one-dimensional system.

The longitudinal confinement in the considered system (\ref{ham}) is taken in the form of an attractive Gaussian potential of the depth $V_0$, which is commonly used to model quantum dots~\cite{xie,mo,Adam}. The two-particle Hamiltonian (\ref{ham}) spectrum is continuous above the threshold energy $\varepsilon_{th}^{(2)}=\varepsilon^{(1)}$, where $\varepsilon^{(1)}$ is the one-particle energy. The bound-states are associated with the solutions of the Schr\"{o }dinger equation under vanishing boundary conditions, the eigenenergies of which are real and less than $\varepsilon_{th}^{(2)}$. The autoionizing resonances correspond to the discrete solutions that satisfy outgoing boundary conditions. The resonance eigenvalues are complex numbers
\begin{equation}
E = \varepsilon - i\frac{\Gamma}{2},
\label{complex}
\end{equation}
which determine the binding energy $\varepsilon$ and the inverse of the resonance lifetime $\Gamma$. In the present work we investigate both singlet (spatially symmetric) and triplet (spatially antisymmetric) states.

\section{The method}
Since the eigenvalue problem of the Hamiltonian (\ref{ham}) does not admit analytical solutions, calculations must be  performed numerically. In order to obtain both the bound and resonant states, we apply the configuration interaction (CI) expansion 
\begin{equation} \Psi_{s,t}(x_1,x_2)=\sum_{i,j} a_{ij} \psi_{ij}^{\pm}(x_1,x_2), \end{equation}
where 
\begin{equation}\psi_{ij}^{\pm}(x_1,x_2)=c_{ij}\left(\phi_i(x_1)\phi_j(x_2)\pm\phi_j(x_1)\phi_i(x_2)\right), \end{equation}
where
$$c_{ij}=\left\{
\begin{array}{ccc}
\frac{1}{\sqrt{2}}&\mbox{}&i\neq j \\
\frac{1}{2}&\mbox{}&i= j \\
\end{array}
\right. ,$$ 
which ensures the proper symmetry under permutations of the particles, so that $(+)$ and $(-)$ correspond to the singlet (s) and
triplet (t) states, respectively. Here we choose the single particle orbitals as the harmonic oscillator (HO) eigenfunctions 
\begin{equation}  \phi_j(x)=\left(\frac{1}{\sqrt{\pi}2^j j!}\right)^{1/2} H_j (x) e^{-\frac{x^2}{2}}, \label{baza}\end{equation}
where the functions $H_i (x)$ are the Hermite polynomials. The whole spectrum of the system is determined by exact diagonalization of the infinite Hamiltonian matrix, the elements of which are given by \begin{equation}H_{nmij}=\int_{-\infty}^{\infty}\psi_{nm}(x_1,x_2)\hat{H}\psi_{ij}(x_1,x_2)dx_1dx_2. \label{eq:real}
\end{equation}
Diagonalization of truncated matrices $[H]_{M\times M}$ yields $M$th
order approximations to wave functions and the corresponding energies of $M$ states. The
accuracy of the method can be systematically improved by
increasing the number $M$ of basis functions, obtaining successive
approximations to the larger and larger number of states. 
In the strictly one-dimensional limit of $\delta\rightarrow 0$ the direct calculation for symmetric wave function become divergent. Fortunately, the ground state of the strictly one-dimensional interacting system ($g\ne 0$) can be determined avoiding divergences by mapping its wave function onto the lowest energy antisymmetric wave function $\psi_{F}$ via the Bose-Fermi mapping relation $\psi(x_{1},x_{2})=|\psi_{F}(x_{1},x_{2})|$~\cite{astr}.

The CI method can be generalised to determine resonant states by using the complex scaling transformation $U(\theta): x \mapsto x e^{i\theta}$. The spectrum of the complex-rotated Hamiltonian 
\begin{equation} \hat{H}_{\theta}=U(\theta) \hat{H}U^{-1}(\theta)  \end{equation}
is described by the Balslev-Combes theorem \cite{ba}, which states that the real bound-state eigenvalues, the complex resonance eigenvalues and the thresholds are the same as those of the original Hamiltonian, but the eigenvalues of the continuous spectrum are rotated about the thresholds by an angle $2\theta$ into the lower energy half-plane, exposing complex resonance eigenvalues. The theorem is proven for dilatation analytic potentials~\cite{ba,NHQM}. However, the application of the complex scaled CI method for potentials that do not have this property~\cite{yaris,Corcoran,nmmp,Buenker} appeared successful and it has been argued~\cite{MorganSimon} that such an approach can be viewed as finite matrix approximation to the mathematically precise exterior complex scaling~\cite{Simon}. 

\begin{figure}[h]
\begin{center}
\includegraphics[width=0.65\textwidth]{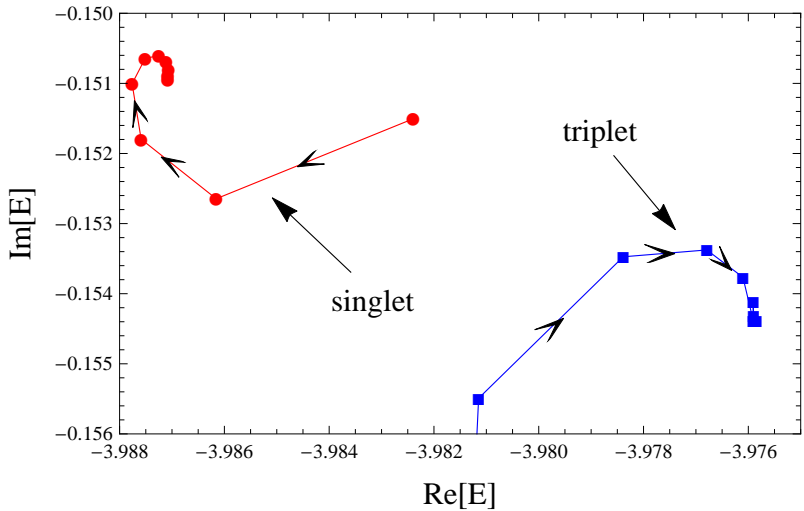}
\end{center}
\caption{The $\theta$ trajectories for two lowest energy resonances for exemplary values of the system parameters $g=8$, $V_0=10$, $\delta=0.01$. Different points represent different values of $\theta$ from 0.1 to 0.75 in steps of 0.05 radians.}

\label{theta}
\end{figure}

Based on this findings, we apply the complex scaled CI method to the model system (\ref{ham}), where the soft Coulombic potential~(\ref{pot}) is non dilatation analytic. We determine the eigenstates of the system through diagonalization of the truncated Hamiltonian matrix $[H]^{\eta}_{M\times M}$, the elements of which are obtained as
\begin{equation}
H^{\eta}_{nmij}=\int_{-\infty}^{\infty}\psi_{nm}(x_1\eta ,x_2 \eta )\hat{H}\psi_{ij}(x_1\eta ,x_2\eta )dx.\label{scale}
\end{equation}
The Hamiltonian matrix elements (\ref{scale}) are analytical function of $\eta$ and therefore we can analytically continue them to the complex plane by
substituting $\eta=e^{-i \theta}$ as first proposed by Moiseyev and Corcoran \cite{Corcoran}. The resonance eigenvalues are determined through the stabilization procedure \cite{mo} as stationary solutions in the complex space
\begin{equation}\frac{d E_k^{\theta}}{d \theta} \Big|_{\theta=\theta_{opt}}=0. \label{omegaopty}
\end{equation}
The Fig. \ref{theta} shows how the exemplary solutions of (\ref{omegaopty}) are connected with a cusp in $\theta$ trajectories in the complex energy plane.

\section{Results: energy and lifetime}
First, we study how the energy spectrum of the two-particle Hamiltonian~(\ref{ham}) depends on the longitudinal potential, the lateral confinement range and on the interaction strength $g$. The calculations were performed with the number of basis functions $M=342$ in the singlet case and $M=324$ in the triplet case, which proved sufficient to obtain convergent results.

\subsection{Dependence on the longitudinal confinement depth}
\begin{figure}[h]
\begin{center}
\includegraphics[width=0.65\textwidth]{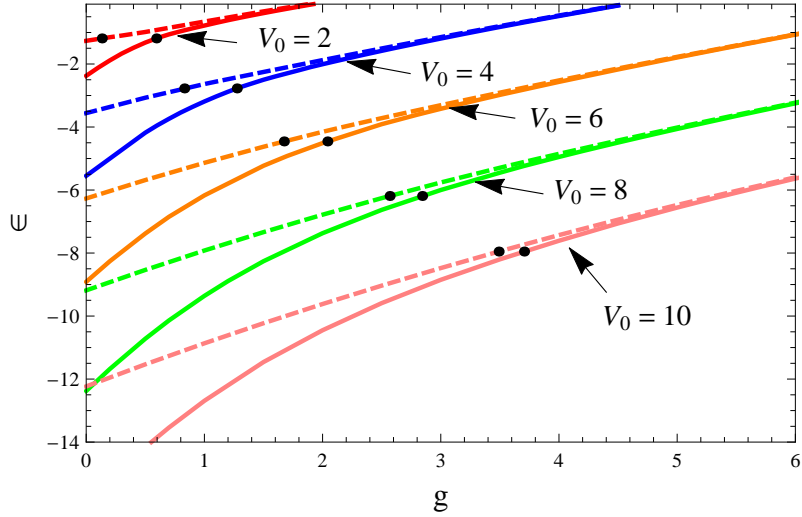}
\end{center}
\caption{The energies of the singlet (solid curve) and triplet (dashed curve) states as functions of the interaction strength $g$ for different depths $V_0$ of the trap. The black points represent the thresholds $g_{th}$ which separate the bound states from resonances.}
\label{ener}
\end{figure}

The analysis of the effect of the depth of the longitudinal trapping potential on the spectrum of the Hamiltonian (\ref{ham}) will be performed at fixed lateral confinement range with the related parameter $\delta$ set to $0.01$. In Fig. \ref{ener} the energies of the lowest singlet and triplet state are presented as functions of the interaction strength $g$ at five different depths $V_{0}$ of the trapping potential. 
As one can see, the depth of the trap has an important effect on the critical value of the interaction strength $g_{th}$ at which the bound state is transformed into a resonance, namely the larger is the value of $V_0$, the larger is $g_{th}$. Generally,
the energies of singlet states lie below the corresponding triplet ones and the singlet-triplet degeneracy is achieved in the limit of $g\rightarrow\infty$. The dependence of the triplet energies on $g$ is much weaker and the thresholds are lower than the ones of the singlet states.

\begin{figure}[h]
\begin{center}
\includegraphics[width=0.65\textwidth]{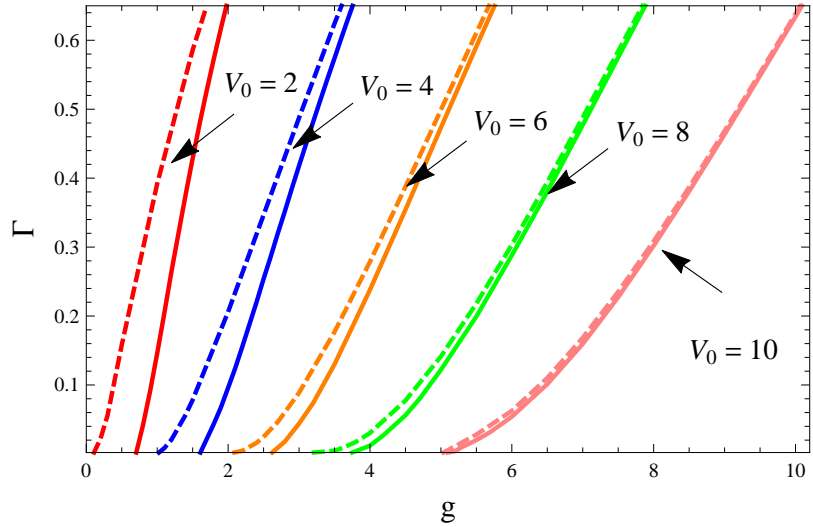}
\end{center}
\caption{The widths of the resonant singlet (solid curve) and triplet (dashed curve) states as functions of the interaction strength $g$  for the same values of $V_{0}$ as in Fig. \ref{ener}.}
\label{gamma}
\end{figure}

Above the autoionization thresholds, the energy eigenvalues acquire an imaginary part which determines the width $\Gamma$ of the corresponding resonance state. We observe in Fig. \ref{gamma} that the widths of resonant states are monotonically increasing functions of $g$ that start at the thresholds $g_{th}$. 
The slope of the functions decreases and the singlet and triplet curves approach each other when the depth of the Gaussian trap $V_0$ increases. This means that the lifetime of resonant states increases with increasing $V_0$ and decreasing $g$. The singlets decay faster than the corresponding triplets, but the differences diminish with increasing depth of the trap.

\subsection{Dependence on the lateral radius}
The influence of the lateral confinement range on the energy spectrum will be studied by varying the $\delta$ parameter for a trap of fixed depth $V_0=10$. 
\begin{figure}[h]
\begin{center}
\includegraphics[width=0.5\textwidth]{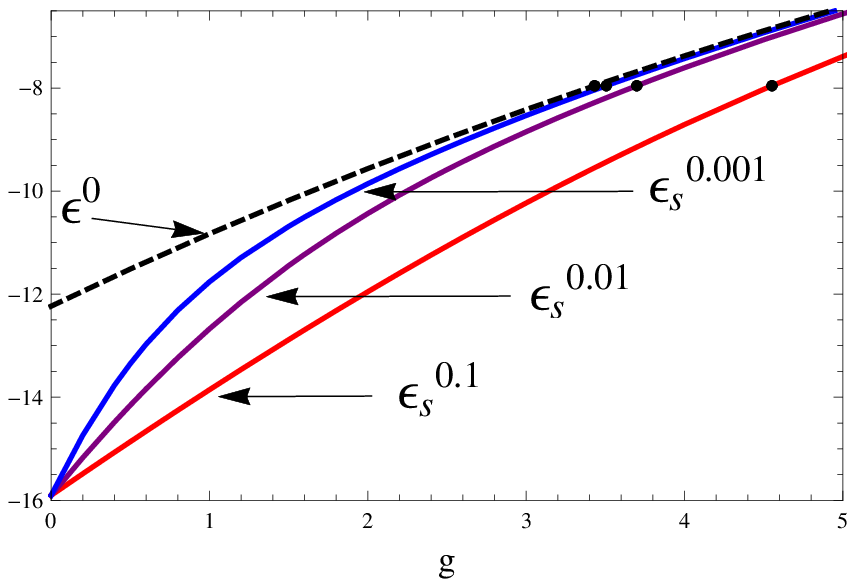}\includegraphics[width=0.5\textwidth]{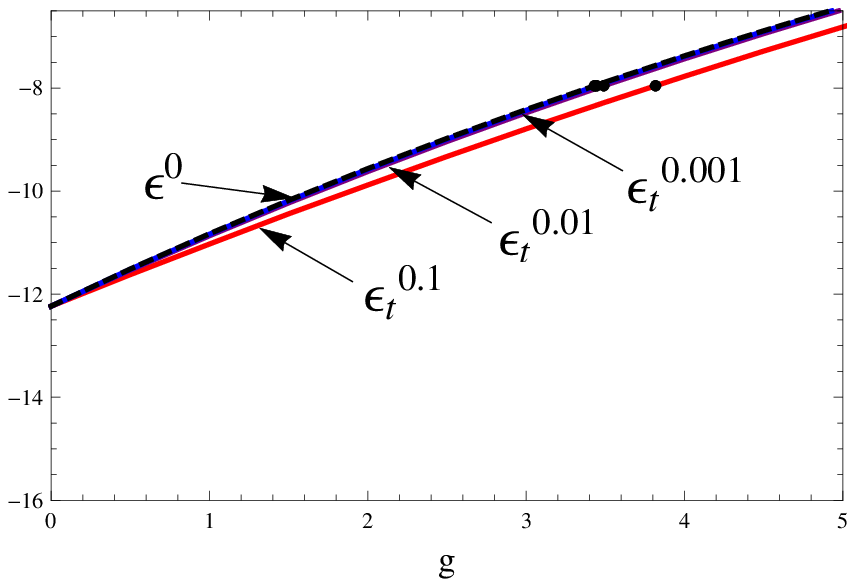}
\end{center}
\caption{The energies of the singlet (left) and triplet (right) states in the trap of depth $V_0=10$ for different values of $\delta$. The thresholds are marked by black points.}
\label{enst}
\end{figure}
In~Fig.~\ref{enst}, the energies of the singlet and triplet states are presented as functions of the interaction strenght $g$. For the pure Coulomb interaction $\delta=0$, the singlet and triplet energies are degenerate, except at the point $g=0$ where the singlet curve is discontinuous. Both the singlet and triplet energies monotonically decrease when $\delta$ increases. However, the behavior of the curves in the vicinity of $g=0$ is markedly different, the triplet one approaches the continuous pure Coulomb curve, reaching the value about $-12.2$ at $g=0$, while the singlet one tends to the discontinuous pure Coulomb curve, reaching the value about $-15.9$ at $g=0$. The threshold values of the interaction strength $g_{th}^{\delta}$, which separate bound states from resonances, being ($g_{th}^{0.1}~\approx~4.55$, $g_{th}^{0.01}~\approx~3.7$, $g_{th}^{0.001}~\approx~3.51$, $g_{th}^{0}~\approx~3.43$) in the singlet case,
and ($g_{th}^{0.1}~\approx~3.82$, $g_{th}^{0.01}~\approx~3.49$, $g_{th}^{0.001}~\approx~3.44$, $g_{th}^{0}~\approx~3.43$) in the triplet case are marked Fig. \ref{enst}. In both cases the thresholds get smaller with decreasing~$\delta$.

\begin{figure}[h]
\begin{center}
\includegraphics[width=0.49\textwidth]{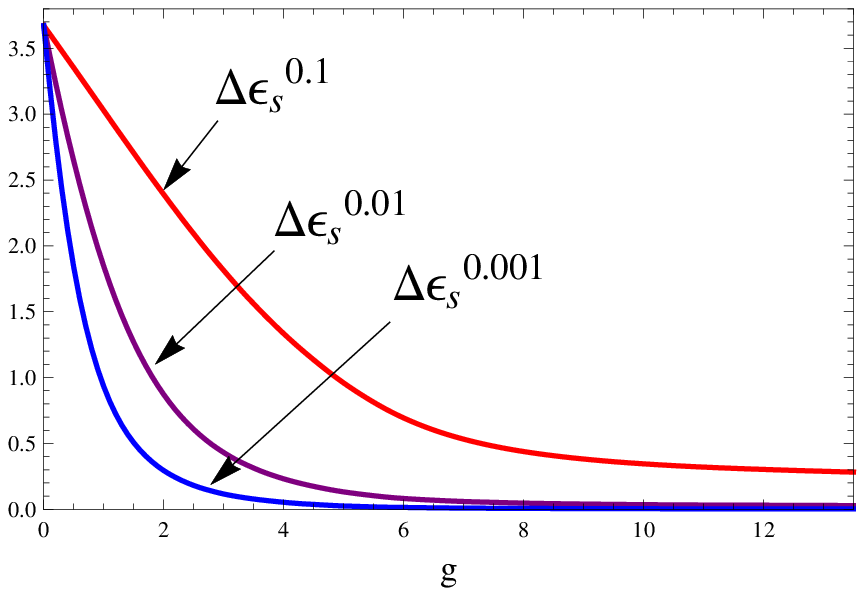}
\includegraphics[width=0.49\textwidth]{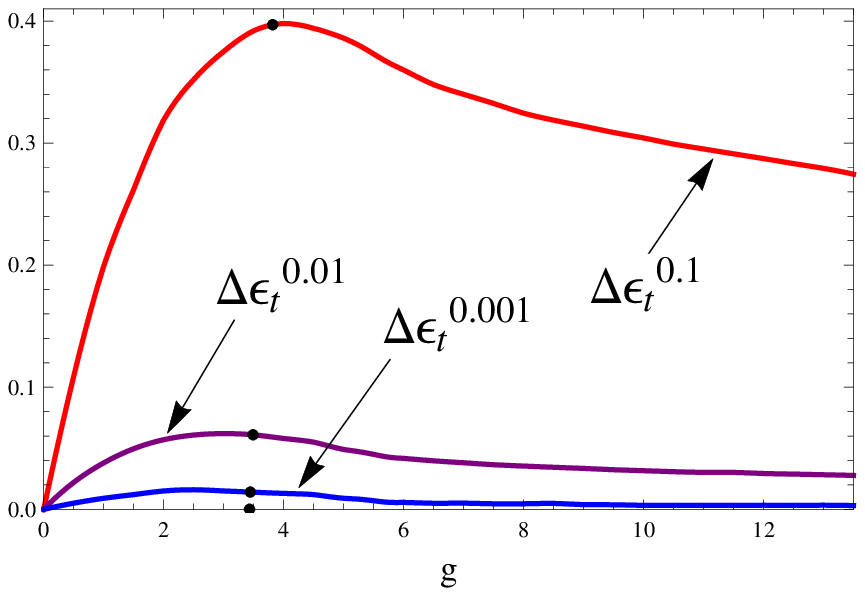}
\end{center}
\caption{The energy differences $\Delta \varepsilon_{s,t}^{\delta}$ of the singlet (left) and triplet (right) states in the trap of depth $V_0=10$ for different values of $\delta$.}
\label{delta1}
\end{figure}

\begin{figure}[h]
\begin{center}
\includegraphics[width=0.65\textwidth]{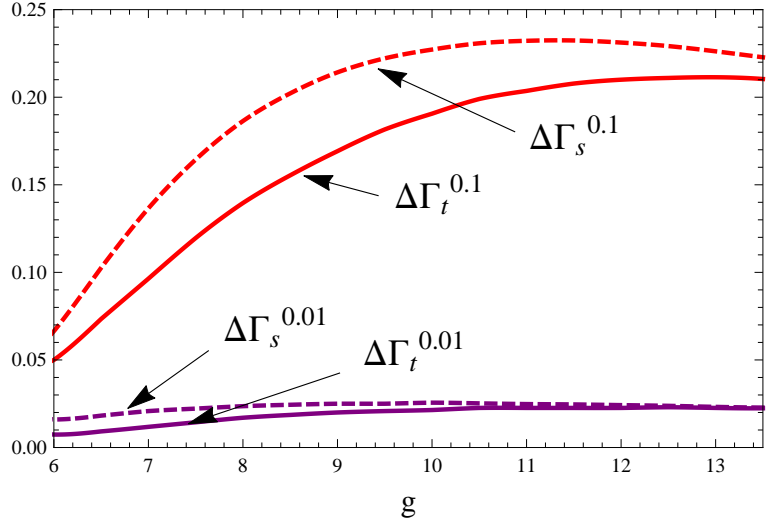}
\end{center}
\caption{The width differences $\Delta \Gamma_{s,t}^{\delta}$ of the singlet (dashed lines) and triplet (solid lines) states in the trap of depth $V_0=10$. }
\label{delta3}
\end{figure}

In order to examine more closely the dependence on the lateral confinement, the system with effective interaction (\ref{pot}) at a given value of the parameter $\delta$ will be compared with the purely Coulombically interacting system. For the singlet (s) and triplet (t) states, we define the energy differences 
\begin{equation}
\Delta \varepsilon_{s,t}^{\delta}=\varepsilon^{0}-\varepsilon_{s,t}^{\delta},
\end{equation}
where the value of $\varepsilon^0$ is obtained for pure Coulomb interaction. In Fig. \ref{delta1} we can see that $\Delta \varepsilon_{s,t}^{0.001} \le \Delta \varepsilon_{s,t}^{0.01} \le \Delta \varepsilon_{s,t}^{0.1}$ irrespective of the interaction strength. For singlet state, we observe a significant influence of the parameter $\delta$ on the energy differences in the vicinity of $g=0$, which is related to the discontinuity of the energy curve at this point in the case of pure Coulomb interaction. Interestingly enough, in the triplet case we observe that the energy differences are maximal in the vicinity of autoionization thresholds, which are marked by dots in Fig. \ref{delta1}. In both the singlet and triplet cases, the energy differences $\Delta \varepsilon_{s,t}^{\delta}$ decrease at large $g$, but the decrease rates are smaller for larger $\delta$. 
We observed that the widths of the singlet and triplet resonances  are monotonically increasing functions of the interaction strength $g$. In both cases they approach the pure Coulomb interaction curve from below when $\delta$ parameter decreases. This means that the resonance lifetimes are the shortest in the case of strictly one-dimensional systems. In Fig. \ref{delta3}, we show the differences of the resonance widths
\begin{equation}
\Delta \Gamma_{s,t}^{\delta}=\Gamma^{0}-\Gamma_{s,t}^{\delta},
\end{equation}
where $\Gamma^0$ is obtained for pure Coulomb interaction. After initially increasing, the differences $\Delta \Gamma_{s,t}^{\delta}$ go through the maxima and then slowly decrease with increasing $g$. For smaller $\delta$ the differences from the case of pure Coulomb interaction are smaller, being invisible in the scale of Fig. \ref{delta3} already for $ \delta~=~0.001$.

\section{Conclusion}
The stability properties of the quasi-one dimensional two-electron quantum dot depend strongly on the shape of the confining potential and the interaction strength $g$. At fixed lateral confinement range, the critical value of the interaction strength $g_{th}$ at which the bound state is transformed into a resonance, increases with the depth of the longitudinal potential $V_0$. The energies of singlet states lie below the triplet ones, becoming equal in the limit $g\rightarrow\infty$. The lifetime of resonant states increases with increasing $V_0$ and decreasing $g$. 

The lateral confinement range also influences the energies and the stability properties. For the strictly 1D system, the singlet and triplet energies are degenerate, except at the point $g=0$ where the singlet curve is discontinuous. When the lateral radius increases, both the singlet and triplet energies monotonically decrease. At small values of $g$, its influence for singlets is much stronger than for triplets. For triplet states, the dependence on the lateral radius is the most visible near the ionization thresholds. Whereas, the resonance lifetimes of singlets and triplets monotonically increase with increasing lateral radius of the quantum dot.

\begin{acknowledgements}
We would like to thank Dr Przemys\l{}aw Ko\'scik for helpful comments and critical reading of this manuscript. We are also grateful to the referee for valuable remarks and suggestions.
\end{acknowledgements}



\end{document}